\documentclass[useAMS, usegraphicx,usenatbib]{mn2e}

\title{The power spectrum of the residual  rotation curve velocity  as a probe of past mergers}
\author[Goldman]
       {Itzhak Goldman\thanks{I thank  the Department of Astronomy and Astrophysics, Tel Aviv University, for
the hospitality while on Sabbatical Leave from Afeka College.}  \\
  Department of Exact Sciences, Afeka  Tel Aviv Academic College  of  Engineering,
218 Bnei Efraim St., Tel Aviv 69107, Israel\\  email:
goldman@afeka.ac.il }

\begin{document}

 \date{Accepted .
      Received ;
      in original form }

\pagerange{\pageref{firstpage}--\pageref{lastpage}}
\pubyear{2010}

\maketitle
\label{firstpage}

\begin{abstract}  
According to the $\Lambda CDM$ cosmological framework, galaxies underwent multiple mergers in their history.
In this   paper we propose to use the power spectrum of the residual fluctuations of the rotation curve velocity as a probe of past mergers. The proposition relies on the assertion that mergers are expected to induce large scale flows and in case of major mergers   shocks are induced as well. Instabilities of the large scale flows and shocks could generate
a large scale turbulence whose size is comparable to the galactic disk dimensions.
We develop expressions relating underlying
turbulence spectral function to the observational power spectrum of the residual of the rotation curve velocity.  This relation can be used  to test whether   turbulence exists  in a given galaxy. The method is applied to the regular spiral galaxy NGC3198 with the conclusion that
it underwent a minor merger about 7 Gyr ago.

\end{abstract}
 \begin{keywords}
 galaxies: general,   galaxies: kinematics and dynamics, turbulence,ISM: kinematics and dynamics, galaxies: mergers, galaxies: individual (NGC3198) 
\end{keywords}

\section{Introduction}
 
The standard $\Lambda CDM$ cosmology implies that galaxies grow during their history by accretion and mergers \citep{{Robertson}, {Kaviraj},{Stewart}}. 
We propose to use the power spectrum of the residual fluctuations of the rotation curve velocity  as a probe of past mergers. The proposition relies on the plausible assumption  that mergers  induce large scale flows, and in the case of major mergers   shocks are induced as well. The large scale flows and shocks can   generate   a turbulence, in the ISM. The physical parameters of the ISM make the generation of turbulence quite easy once an instability and an energy source are available and indeed turbulence in the ISM  is quite ubiquitous \citep{ElmegreenScalo}. The instabilities related to mergers are
  e.g. the  Kelvin Helmholtz instability and  
 the Richtmyer-Meshkov instability  
  due to the shock acceleration    \citep{{Mikaelian},
 {Graham}}.
 
 The spatial scale of the generated turbulence is expected to be comparable to the galactic disk size.  
Such a large scale  turbulence, is different from local turbulence generated by e.g.  supernovae (scale of the order of 10 \ pc), super shells resulting from chains of supernovae (scale of the order of 100 \ pc), or winds from early type stars (scale of the order of $10-100$ \ pc). Moreover, a spatial  scale comparable to a galactic size implies  for turbulent rms velocities of the order of $10 km/s$, a turbulence life time of a few Gyr. For minor  to medium mergers (the   most common)   the turbulence is expected to 
be subsonic and so its life time is even  longer. 
The  long life time of the    turbulence implies that it can be detected also in nearby galaxies. 
 Thus, the turbulence serves as a fossil evidence of the past merger.

 If such  primordial turbulence exists, it is expected to 
  be superimposed on the rotation velocity. Hence, part (or all) of the fluctuating residual   of the observational rotation velocity may be identified with this primordial turbulence. Hints for the existence of such a turbulence are found in observations of   radial velocities in disks of galaxies \citep{{Beau},{wong},{Elson},{Trachter}}  and the observation of  \cite{Begeman}
that the residual of the rotation velocity as function of the azimuthal angle along a ring in the galaxy plan, is oscillatory.  The above velocities are of the order of $5-10 km/s$.  

The theoretical consideration and the observational hints  motivate one  to compute the observational  power spectrum  
of the fluctuating residuals of the rotation velocity. The  residuals depends of course on the definition of the  mean  rotation curve.  To avoid this ambiguity we shall consider only galaxies with a flat   rotation curve over  a sizable part of the radial extent. The residuals are defined with respect to the constant mean value of the rotation velocity.
The resulting observational power spectrum can be used to distinguish between  random residuals (or local small size fluctuations) and   fluctuations which are
a manifestation of a galactic scale turbulence. If the latter  is revealed, it means that it has been generated by a process with a spatial coherence scale comparable to that of the galaxy. A merger or close passage by another galaxy   is the natural candidate. 

  We derive the relation  between the power spectrum of the observational residual rotation velocity, and the underlying 3D turbulence spectrum. This relation can be applied to determine   the turbulence 3D spectral function  from the observational power spectrum. In addition,  
the latter supplies an estimate for  the line of sight depth  of the turbulent region.  

The method is demonstrated for the rotation curve of the spiral galaxy NGC3198 and leads to the conclusion that in spite its ordered shape, this galaxy  has experienced a minor merger about 7 Gyr ago.

\section{ power spectrum of the residual rotation velocity curves}

To test whether  a large scale turbulence  exists in a given galaxy one needs   the relation between  
  the power spectrum of the observational residuals   and the spectral function of the underlying 3D turbulence. The observational  rotation curve is usually obtained    by use of the tilted ring model  \citep{Rogstad74}. We show that its power spectrum gives essentially the same results as a power spectrum of the residual Position -- Velocity (PV) curve along a given axis on the galactic disk. The advantage   is  that the rotation velocity at a given radius,   is based on many measurements at this radius, along the ring circumference and so the precision is better than that of the PV data which is based on a single measurement for each radius..

\subsection{Turbulence Spectrum }
 
The  3D spectral function of the turbulent velocity, $\Phi(\vec{k})$ is defined in terms of the 2-point
correlation  of the turbulent velocity field $<\vec{v}(\vec{r'}) \cdot \vec{v}(\vec{r}+\vec{r'})> $  where the angular brackets denote ensemble averaging  \citep{Lesieurs}. In practice, the ergodic assumption is invoked and the ensemble average is replaced by space, area or line averages. For homogeneous turbulence the two point correlation is  a function of the  separation  between the two points, so that

\begin{equation}
\label{phi}
\Phi(\vec{k}) = \frac{1}{(2 \pi)^3}\int <\vec{v}(\vec{r'}) \cdot \vec{v}(\vec{r}+\vec{r'})>   e^{i \vec{k}\cdot \vec{r}}  d^3r 
\end{equation} 
 the inverse transform yields 
\begin{equation}
\label{phi1}
 <\vec{v}(\vec{r '}) \cdot \vec{v}(\vec{r}+\vec{r'})>    =  \int \Phi(\vec{k})  
  e^{-i \vec{k}\cdot \vec{r}}  d^3k 
\end{equation} 
In  the homogeneous and isotropic case $\Phi$ depends only on the absolute value of the wave number and  it is useful to introduce the
turbulence energy spectrum $E(k)$ and the turbulent velocity spectral function $F(k) = 2 E(k)$ so that

\begin{equation}
\label{F}
 \Phi(\vec{k})=\Phi(k)=\frac{  F(k)}{4\pi k^2}\ ;  k=| \vec k| 
 \end{equation}

The measured radial velocity at each position in the plane of the sky, is an intensity-weighted average of the velocity contributed by  the emitting gas  along the line of sight. A simplifying assumption of homogeneity translates this into a line of sight average of the velocity.

In the following two subsections we assume that a 3D large scale turbulence exists and derive the relations between the   observational power spectrum of the PV and rotation curve data
and the underlying 3D turbulence. The simplifying assumptions of homogeneity and isotropy are used.

 \subsection {The Power Spectrum of Position Velocity Data}
 
 Let us consider first  the power spectrum of a Position-Velocity (PV) data, namely the rotation velocity as function of position along a given axis of the galaxy disk. This power spectrum is  defined as Fourier transform of the two point correlation of the observed residual  velocity.   Denoting the axis as $y=0$ and the line of sight as $z$, and the residual observed rotation velocity as $u$, the power spectrum is

\begin{equation}
\label{ppv1}
 P_{pv}(q) =\frac{1}{2\pi} \int e^{ i q x}   <u(x +x', 0, z) u(x', 0,z')> dz dz'  dx 
 \end{equation} 
 Using the isotropy assumption  and Eq. \ref{phi1} one finds
 
 \begin{equation}
\label{corr}
  <u(x +x', 0, z) u(x', 0,z')>=\frac{1}{3} \int \Phi(\vec{k})   e^{-i ( k_x x + k_z(z-z')) }  d^3k  
 \end{equation} 
 which when inserted in Eq. \ref{ppv1} leads to
 \begin{equation}
 \label{ppv2}
 P_{pv}(q)=  \frac{1}{3}  \int \Phi(q,k_y, k_z) e^{   i k_z (z-z')}      dk_y dk_z dz dz' 
\end{equation}
 and   Integrations over $z, z'$ yield
  \begin{equation}
 \label{ppv3}
  P_{pv}(q, D)=  \frac{1}{3}  \int \Phi(q,k_y, k_z)   \left(\frac{\sin(k_z D   /2)}{(k_z D /2)}\right)^2dk_y dk_z  
 \end{equation}
 with $D$,  the depth along the line of sight of the turbulent region. In the case of a disk galaxy $D \cos i = 2 H$ with $H$ 
 denoting the   scale height of the disk and $i$  the inclination angle.

For an  homogeneous and isotropic turbulence, the 3D turbulence spectral function  is a power law of the absolute value of the wavenumber: $\Phi(q, k_y, k_z)= A (q^2 + k_y^2 + k_z^2) ^{-m}$.

 In the incompressible subsonic case   the spectrum
 is the Kolmogorov spectrum with  $m=11/6$  \citep{Lesieurs}. For compressible supersonic turbulence
 $m=2$  \citep{{Passot}, {GirimajiZhou}}. The steeper slope
is due to the fact that a fraction of the turbulent kinetic energy density at a given wavenumber is converted to compression work  decreasing the rate of energy transfer to the larger wavenumbers.

Integration over $k_y$ from $-\infty$ to $\infty$ yields
$$  P_{pv}(q, D) =B(m)\int_{-\infty}^{\infty}(q^2  + k_z^2) ^{(-m+1/2)}\left(\frac{\sin(k_z D   /2)}{(k_z D /2)}\right)^2 dk_z$$ 
where $B(m)$ is a constant depending on   $m$ and $D$.  One can further express the power spectrum in the form
 \begin{equation}
 \label{ppv}
  P_{pv}(q, D) =B(m)\int_{-\infty}^{\infty}(   q D/2 )^2  +\eta^2)^{(-m+1/2)}\  \frac{\sin^2\eta}{\eta^2}  d\eta  
 \end{equation} 
The power spectrum  is a power law with index $2 -2 m$ for $ q D/2<<1$ and $1 -2 m$ 
 for $ q D/2>>1$. This effect of the change of the power law index was
found observationally by \cite{Elmegreen2001} for the $H_I$ power spectrum  of the LMC, by \cite{Dutta} for the $H_I$ power spectrum of NGC1058 and recently by  \cite{Block} for the infrared power spectrum of the LMC, and by \cite{Contini} for both the velocity and mid infrared power spectra of the shocked  nebulae near the turbulent Galactic Center.  The results of the last two references are based on the Spitzer data.
 
\subsection{The Power Spectrum of the Rotation Curve}

The  rotation curve is obtained by use of the tilted ring model  \citep{Rogstad74} in which the disk is approximated as a superposition of concentric rings or annuli. The observed velocity   field is then used to derive  for each ring a rotation velocity, an inclination angle and a position angle.
 The rotation curve velocity as function of radius obtained in this way is more accurate than 
the PV curve, since the value at each radius is derived from multiple observational values.

This motivates a computeation of the power spectrum of the residual fluctuations of the rotation curve and its  use to test for the existence of an underlying 3D turbulence. For random fluctuations the power spectrum is expected to be independent of the wavenumber. If however,
the power spectrum varies in an ordered manner with the wavenumber it is possible that it
reflects an underlying turbulence.  

As in the case of the PV power spectrum there is the need to obtain the relation between the observationally computed power spectrum and the underlying turbulence spectral function. In the case of the rotation curve this task seems a prioiri more difficult than the  PV case since the rotation velocity is fitted to the observed line of sight velcities along the ring circumference
and not measured durectly along a given line in the plane of the disk. For simplicity the fit along the ring is represented as a simple average over its circumference, so that
 \begin{equation}
\label{prot1}
  P(q) =\frac{1}{(2\pi)^3} \int e^{i q r}<u(r +r',\theta, z) u(r', \theta',z')> d\theta d\theta' dzdz'
  \end{equation}
 As in the PV case, here too use of Eq. \ref{phi1} and assuming  isotropy of the turbulence imply
\begin{equation}
  \label{prot2}
P(q) =\frac{1}{3} \frac{1}{( 2\pi)^3}  \int \Phi(k, k_z)  e^{-i k( r+ r')\cos(\alpha -\theta )}  e^{i k r'  \cos(\alpha -\theta' )} 
   \end{equation} 
 
$$e^{ i q r}e^{-i k_z (z - z')}  dr d\alpha k dk dk_zdz dz'  d\theta d\theta' $$
  where  $k = \sqrt{k_x^2 + k_y^2}$ is the wavenuber in the plane of the sky and $\alpha$ denotes its azimuthal angle. In the integration along the ring circumference $\alpha$ spans an interval of $2 \pi$ and so do $\alpha -\theta$ and $\alpha -\theta'$. One gets,

 \begin{equation}
 \label{prot3}
P(q) =\frac{1}{3} \frac{1}{( 2\pi)^3}  \int \Phi(k, k_z)e^{ i q r}e^{i k_z (z - z')}
\end{equation}
$$ e^{-i k  r cos \beta}     dr d\beta k dk dk_zdz dz'  d\theta d\theta'  = $$
 $$\frac{1}{3}    \int \Phi(k, k_z)e^{ i q r}  J_0( k r) k\left( \frac{\sin(k_z D   /2)}{(k_z D /2)}\right)^2 dk dr dk_z  $$
  with $J_0$ denoting the zero order Bessel function of the first kind.
 
For an underlying 3D turbulence characterized by a power law spectral function  $\Phi(k, k_z)= A (k^2   + k_z^2) ^{-m}$,  the integrations on $ r$  and $k$ yield 
 \begin{equation}
\label{prot4}
 P (q) =B_1(m)\int_{-\infty}^{\infty}(q^2  + k_z^2) ^{(-m+1/2)}\left(\frac{\sin(k_z D   /2)}{(k_z D /2)}\right)^2 dk_z 
   \end{equation}
 where $B_1(m)$ is a constant depending on the index $m$ and $D$ is the depth along the line of sight. The analytic integrations over r and k leading to Eq. (\ref{prot4}) were done using the \cite{mathematica} software. One can further express the power spectrum in the form
 \begin{equation}
  \label{prot} 
  P  (q) =B_1(m)\int_{-\infty}^{\infty}(   q D/2 )^2  +\eta^2)^{(-m+1/2)}\  \frac{\sin^2\eta}{\eta^2}  d\eta  
   \end{equation}
Note that the observational power spectrum of the residual velocity curve  has the same relation with respect to the underlying 3D turbulence as  does  the observational PV power spectrum.  This result should not come as a surprise since the rotation curve can be regarded as an  averaging of  PV data along many axes. As a result, the power spectrum of its residual fluctuations is an   average of the   PV power spectra and hence share the same relation  with the spectrum of the underlying 3D turbulence. As noted above, it is expected to be more accurate than the PV power spectrum.

\begin{figure} 
  \centerline{\includegraphics[scale=1 ]{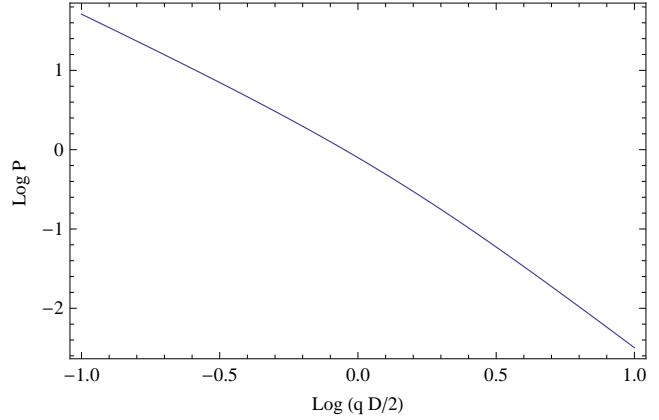}}
\caption{$P(q)$ in dimensionless units, as function of $q D/2$ for $m=11/6$} 
\label{fig:pq} 
\end{figure}

Fig.  \ref{fig:pq}   displays the observational power spectrum as function of $q D/2$ for 
an underlying  3D  Kolmogorov turbulence characterized by  $m=11/6$.  
 For spatial scales much larger than the depth along the line of sight,   the index of the power spectrum is $-5/3$ and for spatial scales much smaller than the depth it approaches $-8/3$.
 The transition between the two regimes is at $k D/2 \simeq 1.4$ and there the logarithmic slope is $-13/6$.

\subsection{The Observational Power Spectrum of the Rotation Curve of NGC9183}
 
The proposed method is demonstrated by applying it to  NGC9183. The latter   appears to be a rather ordered galaxy, although some deviations from regularity were noted \citep{{bosma81},   {vanAlbada}, {Begeman}}. The $H_I$ column densities do not exceed $5\times 10^{20} cm^{-2}$ so the optical depth is rather small. The data presented in Table 2. of \cite{Begeman} show a very flat rotation curve for radial distances in the range $1.5' - 11'$ corresponding to $4.1\   kpc - 29.9\   kpc$.  This is the range of radial distances to which the proposed test is applied.

The residuals (with respect to the mean value)  are plotted in Fig. 2 as function of radial distance in units of arcminutes. Up to 3', the radial distances are spaced by 0.25' and later by 0.5'. Since the computation of power spectra requires even spacings, we omitted the data points at positions $1.75',\ 2.25', \ 2.75'$ thus there remain overall 20 evenly spaced data points. Note that
the uncertainties are   $(2-3)km/s$; substantially smaller than the 15.5 km/s uncertainty in a single measurement \citep{Begeman}. This arises from the use of many measurements along the ring and demonstrates the advantage of using a rotation curve rather than a PV data.

\begin{figure} 
   \centerline{\includegraphics[scale=1]{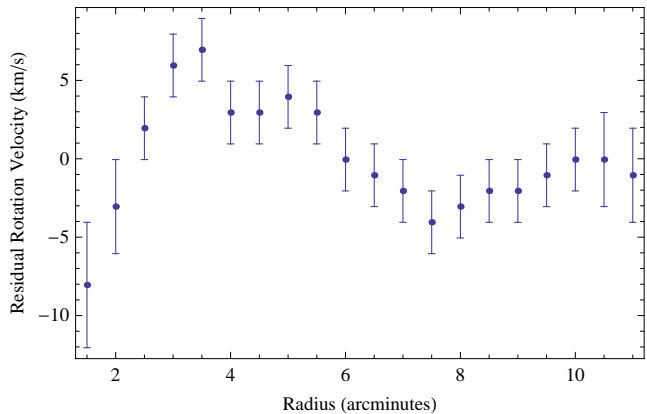}}
\caption{  The residual velocity in units of km/s as function of radial distance in units of arcminutes. The  error bars are taken  from Table 2.   of  Begeman (1989). }   
 \label{vel}
 \end{figure}

The residual velocity    shown in Fig. \ref{vel}
does seem to exhibit  fluctuations, as function of radial distance. But one has yet to test whether these fluctuations do indeed represent a large scale turbulence.
 
 The \cite{mathematica} software was used to compute the power spectrum of the residual rotation curve velocity as function of the wavenumberwas $q= 2\pi/l$, with $l$ denoting the corresponding spatial scale.  Because of the cyclic nature of the discrete Fourier transform, the power spectrum  is obtained for relative wavenumbers   in the range $1-11$ with $q=1$ corresponding $l_0=25.8   kpc$,  the largest spatial scale of the turbulence and $q=11$ corresponds to $2.35 \   kpc$, which is about twice the radial spacing between the observations used in the present analysis.
 
To obtain the error bars for the observational power spectrum we  performed $10^4$ simulations of
``observational''  velocity sets. At each position   a velocity was randomly chosen  from a normal distribution with a mean equal to the observational  value at this position and  a standard deviation equal to the observational uncertainty at this position. 

 For each such set, the power spectrum was computed  and subsequently   the standard deviations, at each wavenumber, of the 
logarithm of the simulated  power spectra.    These standard deviations were taken as the uncertainties in the logarithm of the observational power spectrum of the residual  rotation velocity. 
 
The standard deviation of the observational residual rotation velocity of Fig. \ref{vel}  is $3.6 km/s$. Hence, if it indeed represents   a turbulence the latter must be subsonic(in the isotropic case $v_{turb} = \sqrt{3} 3.6 = 6.2 \ km/s$). The  theoretical fit is thus chosen to be a turbulence spectral function  with $m=11/6$. 
The depth D, along the line of sight  and the overall normalization of the function given by Eq. (\ref{prot}) to minimize the reduced $\chi^2$ value.  

\begin{figure} 
  \centerline{\includegraphics[scale=1]{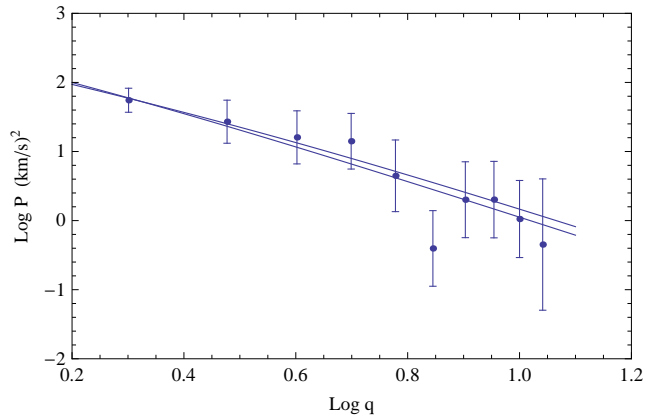}}
\caption{{ \bf Dots}:The logarithm of the power spectrum of the residuals of the observational rotation curve   in units of $(km/s)^2$, as function of the logarithm of the relative wavenumber $q$.   
  {\bf lower curve}: the best fit  $\Phi(q)$: has depth $D=7.31   kpc$ corresponding to an $H_I$ scale height of $1130 \ pc$. {\bf upper curve}: the best fit  $\Phi(q)$  {\it without } the $q=7$ point has depth $D=3.78   kpc$ corresponding to an $H_I$ scale height of $584 \ pc$.} 
 \label{powers}   
 \end{figure}

 The observational power spectrum of the residual rotation curve velocity is shown in Fig. \ref{powers} together with two such fits. The lower one is the best fit with  $\chi^2 = 0.43$. It has a depth along the line of sight $D= 7.31 \   kpc$ translated into a scale height for neutral hydrogen  $H_{H_I} = 1130 \ pc$.  The law value of $\chi^2$ suggests that the uncertainties in    \cite{Begeman} are probably overestimated.
 
 This analysis implies  that the residuals are consistent with being a manifestation of a   subsonic turbulence with a largest scale of $25.8 \ pc$, a turbulent velocity   of $3.6 km/s$, and a lifetime scale of  the order of $7  Gyr$.

  About 75\% of  the $\chi^2 $ value is contributed by the point at relative wavenumber $q=7$ corresponding to a scale of about $3.7 \   kpc$.   
 The upper curve in Fig. \ref{powers} is the best fit  to the power spectrum with this point excluded. It has  a $\chi^2 = 0.11$, and a depth   
  $D=3.78\   kpc$ corresponding to a scale height  $H_{H_I} = 584 \ pc$. 
This latter is more  consistent with  observational values  for disk galaxies \citep{Bagetakos}. 
  For each of the two fits there is a correlation between the normalization of the fit and the value of $D$, resulting in  a flat dependence of $\chi^2$ on D. Thus, in each case  the calculated D and the corresponding scale height vary in a range of about   $\pm 20\%$.
  
\section{Discussion}  

This paper puts forward the idea that analyzing the residual of the rotation curve velocities
could test for the existence of a galactic-scale turbulence. If such a turbulence exits, the large spatial scale implies a lifetime  of the order of a few Gyr. These two features point to a primordial merger  as the natural generating mechanism. The test relies on the fact that
random residuals of the velocity curve would have produce a flat power spectrum in sharp contrast with known turbulence spectra that exhibit a power law dependence on  wavenumber, for wide ranges of the latter. 
 
In this paper the method is proposed for galaxies exhibiting an observational rotation curve which is flat over sizable range of radial distances. This limitation  avoids ambiguity as to the definition of what constitutes the residual velocity curve.  In any case, if the derived observational power spectrum can be identified with   turbulence spectrum, the issue of the definition of the residual is a posteriori solved.

We derived a semi-analytic relation between the observational power spectrum and the spectral function of the underlying 3D turbulence. We proved that the relation is of the same form for the observational spectra calculated from a position velocity data and from the rotation curve. The latter is of course advantageous since at each radial distance it represents a fit along a ring and not a single value.
The method is capable of determining the depth along the line of sight of turbulent gas and hence the disk scale height.

The method has been demonstrated for the rotation curve of NGC3198. The computed power spectrum fits very well a Kolmogorov spectrum with a logarithmic slope of $-5/3$ for the largest scales and changing to $- 8/3$ for the small scales. The turbulent rms velocity is about $6 \ km/s$; the lifetime is of the order of $7 \ Gyr$ and  the  implied scale height  about $580-1100) \  pc$.
For this galaxy, the observationa power spectrum spans a decade in wavenumbers. Clearaly, an higher spatial resolution would have enabled a wider wavenumbre range, inmprove the numerical fit of the the depth and thescale height, and lend more credibility to the cocnclusions. But even so, the case for an underlying 3D turbulence in NGC3198 seems
  quite good.
  
The method can be used as an observational tool for finding how common were mergers. It easy to identify merges in cases of galaxies with a disturbed shape. These mergers are usually major mergers. The majority of the cosmological mergers are expected to be minor mergers that don't have a sharp morphological imprint. The present method enables detecting turbulence also for minor mergers and for galaxies that appear regular, as in the case of NGC3198. The long lifetime of the turbulence enables using it as a fossil evidence for primordial mergers.

It would be interesting to look for turbulence in additional galaxies and specifically, to test for a  correlation between the disk thickness and the turbulent rms velocity.

 \section*{Acknowledgments}  
I thank D. Cheluche and S. Sadeh for discussions, and  H. Goldman for thorough reading and  comments regarding the manuscript. The research was supported by the Afeka College research committee.

  \label{lastpage}
\end{document}